\documentclass[conference]{IEEEtran}
\IEEEoverridecommandlockouts
\usepackage{cite}
\usepackage{amsmath,amssymb,amsfonts}
\usepackage{graphicx}
\usepackage{textcomp}
\usepackage{xcolor}
\usepackage{algorithm}  
\usepackage{algpseudocode}  
\def\BibTeX{{\rm B\kern-.05em{\sc i\kern-.025em b}\kern-.08em
    T\kern-.1667em\lower.7ex\hbox{E}\kern-.125emX}}
\begin{document}

\title{Block Error Performance of NOMA with HARQ-CC in Finite Blocklength
}

\author{
\IEEEauthorblockN{Dileepa Marasinghe, Nandana Rajatheva, Matti Latva-aho}
\IEEEauthorblockA{\textit{Centre for Wireless Communications} \\
\textit{University of Oulu}\\
Oulu, Finland \\
\{dileepa.marasinghe, nandana.rajatheva, matti.latva-aho\}@oulu.fi}
}

\maketitle

\begin{abstract}
This paper investigates the performance of a two-user downlink non-orthogonal multiple access (NOMA) system using hybrid automatic repeat request with chase combining (HARQ-CC) in finite blocklength. First, an analytical framework is developed by deriving closed-form approximations for the individual average block error rate (BLER) of the near and the far user. Based upon that, the performance of NOMA is discussed in comparison to orthogonal multiple access (OMA), which draws the conclusion that NOMA outperforms OMA in terms of user fairness. Further, an algorithm is devised to determine the required blocklength and power allocation coefficients for NOMA that satisfies reliability targets for the users. The required blocklength for NOMA is compared to OMA, which shows NOMA has a lower blocklength requirement in high transmit signal-to-noise ratio (SNR) conditions, leading to lower latency than OMA when reliability requirements in terms of BLER for the two users are in the order of $10^{-5}$.
\end{abstract}


\section{Introduction}
With the advent of new use-cases requiring high reliability and low-latency in 5G and beyond, transmission with finite blocklength becomes inevitable to reduce latency. In contrast to classical information-theoretic principles, the use of finite blocklength results in a non-negligible decoder error probability. Hybrid automatic repeat request (HARQ) procedures are used to improve the accuracy in decoding by exploiting time-diversity at the expense of increased latency. Thus, achieving high reliability and low-latency are Pareto-optimal, which calls for a trade-off between the two. Concurrently, non-orthogonal multiple access (NOMA) has gained widespread attention in research due to the ability to outperform its counterpart, orthogonal multiple access (OMA) in terms of spectral efficiency and user fairness. 

Studies on NOMA with HARQ can be found in the literature \cite{HARQNOMA_choi,choi_harq_IR,Cai:2018,Xu_Outage_and_power}. Authors in \cite{HARQNOMA_choi} show that NOMA with successive interference cancellation (SIC) employing HARQ with incremental redundancy (HARQ-IR) can outperform OMA in outage probability. In \cite{choi_harq_IR}, a power allocation strategy for HARQ-IR with NOMA is presented. The outage performance of NOMA with the HARQ-CC scheme has been studied in \cite{Cai:2018} and \cite{Xu_Outage_and_power}, by deriving closed-form approximations for outage probability. Analysis of HARQ based systems using finite blocklength have been presented in \cite{finiteblockHARQ,greenfiniteblock,Zorzi:2014}. Authors in \cite{finiteblockHARQ} investigate the blocklength which maximizes throughput and minimizes average delay while authors in \cite{greenfiniteblock} investigate power allocation for systems using type-I ARQ in finite blocklength. In \cite{Zorzi:2014}, a closed-form derivation of the outage probabilities on HARQ-IR in finite blocklength is provided. A power allocation method for HARQ-CC with finite blocklength, which targets reliability constraints is proposed in \cite{dosti:2017}. Analysis of NOMA in the finite blocklength regime is reported in \cite{Yu_NOMA_short,Xinyu:2019}. Authors in \cite{Yu_NOMA_short} and \cite{Xinyu:2019} investigate the finite blocklength performance of a two-user downlink NOMA system for single and multiple antenna base station (BS) and demonstrate that having a common blocklength for both NOMA users is optimal and NOMA outperforms OMA in latency.

Improving reliability and minimizing the latency are two targets that are conflicting with each other to be achieved simultaneously. The reason is improving reliability would be supported by re-transmissions including the use of longer packets, which will then increase the latency. A trade-off between latency and reliability that fit different use-cases is required when considering ultra-reliable low-latency communications (URLLC). Some URLLC use-cases are remote surgery and factory automation which have stricter targets like $1\times 10^{-9}$ with 1 ms latency and V2X communications, and tactile internet which have reliability around $1\times 10^{-5}$ and latency requirements ranging from 1 ms to 100 ms. The motivation behind this work is to analyze the performance of a system comprising of the three enablers; NOMA combined with HARQ in finite blocklength. While NOMA allows higher spectral efficiency by utilizing the same frequency-time resource, the use of HARQ improves the reliability, and the use of short packets allows reducing latency. The main goal is to investigate the ability of NOMA to deliver ultra-reliability using HARQ combined with the use of short packets to reduce latency. The reliability is investigated by characterizing the average block error rate (BLER). Also, determining the required number of channel uses or blocklength for NOMA, which satisfies given reliability targets for the two users and comparing the performance of NOMA with OMA are presented . 

The next sections are organized as follows. Section \ref{sec:sysmodel} describes the system model. In Section \ref{sec:NOMA_approx} analytical approximations for the average BLERs are derived for the two users. Section \ref{sec:blocklength_and_power} presents asymptotic BLER approximations considering high SNR conditions and an algorithm is devised to determine the required blocklength for the system to meet given reliability constraints for the two users. In Section \ref{sec:numerical_results} numerical results are provided to validate the derived approximations in Section \ref{sec:NOMA_approx} and comparison of the blocklength requirement between NOMA and OMA is provided. Section \ref{sec:conclusion} concludes the paper. The proofs of the main results are provided in Appendices.

\section{System Model} \label{sec:sysmodel}
Consider a downlink power domain NOMA system that uses short-packets for communications. The system comprises of a single antenna base station and two users $u_1, u_2$ equipped with single antennas. Without loss of generality, assume that $u_1$ is located close to the BS, thus having a higher channel gain, while $u_2$ is located far from the BS with a lower channel gain. Further, assume the channel gain of the users are known at the BS. To enhance the reliability of the transmission, the system uses the HARQ-CC scheme. The BS serves the users following the NOMA principle. Let $x_1$ and $x_2$ be the unit energy messages to $u_1$ and $u_2$, respectively. The BS encodes these messages using the superposition coding technique with power allocation coefficients $\alpha_1$ and $\alpha_2$ such that $\alpha_1 + \alpha_2 = 1$ with a total power of $P$. According to the NOMA principle, BS allocates more power to the far user by setting $\alpha_1 < \alpha_2$ ensuring user fairness. Therefore, the transmitted signal $s$ can be expressed as

\begin{equation}
    s = \sqrt{\alpha_1P}x_1 + \sqrt{\alpha_2P}x_2. \label{eq:message}
\end{equation}

The received signal $y_i$ at $u_i$, $i =  1,2 $ in the $t^{th}$  transmission round can be expressed as

\begin{equation}
    y_i = \tilde h_{i,t}(\sqrt{\alpha_1P}x_1 + \sqrt{\alpha_2P}x_2) + n_i,\label{eq:receivedsignal}
\end{equation}
where $\tilde h_{i,t} = \frac{h_{i,t}}{\sqrt{1+{d_i}^{\eta}}}$,  $h_{i,t} \sim \mathcal{CN}(0,1)$ is the independent and identically distributed (i.i.d) fading coefficient of $u_i$ with equal blocklength $M$ in the $t^{th}$ transmission round, $d_i$ is the distance between $u_i$ and the BS, $\eta$ is the path loss exponent and $n_i$ is the additive white Gaussian noise (AWGN) with variance $\sigma^2$.

The far user, $u_2$ attempts to decode the received signal treating $u_1$'s signal as interference. Then the received signal-to-noise-plus-interference ratio (SINR) at $u_2$ for decoding its message at the $t\textsuperscript{th}$ transmission round is

\begin{equation}
    \gamma^{t}_{22} = \frac{\rho\alpha _2|\tilde h_{2,t}|^2}{\rho\alpha _1|\tilde h_{2,t}|^2+1 }, \label{eq:SNRu22}
\end{equation}
where $\rho$ is the transmit SNR such that $\rho = \frac{P}{\sigma^2}$.

The near user, $u_1$ applies SIC in decoding the messages, which means $u_1$ decodes $u_2$'s message first and then its own message without interference. The SINRs for decoding at $u_1$ are given by
\begin{equation}
    \gamma^{t}_{12} = \frac{\rho\alpha _2|\tilde h_{1,t}|^2}{\rho\alpha _1|\tilde h_{1,t}|^2+1 }   \>\> \textrm{and} \>\> \gamma^{t}_{11} = {\rho\alpha_1|\tilde h_{1,t}|^2}. \label{eq:SNRu11}
\end{equation}

In the HARQ-CC procedure, in case of a failure to decode its message, the user retains the received signal and sends a negative acknowledgement (NACK) to the BS. If a NACK is received to the BS from any of the two users, BS retransmits the same encoded signal. Users employ maximum ratio combining (MRC) for decoding by combining the received signals stored during previous rounds and the new signal received. In case of successful decoding, the user will send a positive acknowledgement (ACK). BS transmits a new signal when it receives ACKs from both users. This work assumes the feedback channel, which ACKs/NACKs are sent, to be a one-bit error-free channel. The number of transmission rounds is limited to a maximum of $T$. The SINR for decoding $u_j$'s signal at $u_i$ where $i,j = 1,2$ after $T$ rounds of transmissions\cite{Cai:2018} is 
\begin{equation}
    \gamma_{ij} = \sum_{t=1}^{T}\gamma^{t}_{ij} .  \label{eq:SNRuij}
\end{equation}

\section{Average BLER of NOMA with HARQ-CC in Finite Blocklength} \label{sec:NOMA_approx}
\subsection{Preliminaries} \label{sec:preliminaries}

Short-packets are used in the system for achieving low-latency in communications with a finite blocklength. Based on the recent work by Polyanskiy et al.\cite{Polyanskiy:2010}, the decoder error probability or the BLER of $u_i$ for decoding $u_j$'s information, $\epsilon_{ij}$ in finite blocklength is given by
\begin{equation}
    \epsilon_{ij} \approx Q \left(\frac{log_2(1+\gamma_{ij})-\frac{N_{j}}{M}}{\sqrt{\frac{v_{ij}}{M}}}  \right) \triangleq \Phi(\gamma_{ij},N_{j},M). \label{eq:decerrrprob}
\end{equation}
where $N_j$ is the number of information bits transferred using a blocklength of $M$ channel uses, $\gamma_{i,j} $ is the SINR, $v_{i,j}$ is the channel dispersion defined by $v_{i,j} = (log_2e)^2 \left( 1- \frac{1}{(1+\gamma_{i,j})^2}\right)$, $Q(\cdot)$ is the Q function. This approximation holds when $M$ is sufficiently large\cite{Xinyu:2019}, such as $M \geq 100$.

The user $u_1$ uses SIC in decoding, so the instantaneous BLER depends on the two stages in the SIC procedure. The success of the first stage affects the BLER in decoding at the second stage. Therefore, the instantaneous BLER for $u_1$ is given by
\begin{equation}
     \epsilon_{1} =  \epsilon_{12} + (1-\epsilon_{12})\epsilon_{11}. \label{eq:eps1}
\end{equation}
Here $\epsilon_{12}$ is the BLER resulting from the first stage of the SIC decoding and $1-\epsilon_{12}$ denotes the success in the first stage. The average BLER $\epsilon_{11}$, results from the interference-free decoding in the second stage. These are respectively given by
\begin{equation}
     \epsilon_{12} = \Phi(\gamma_{12},N_{2},M) \qquad \textrm{and} \qquad \epsilon_{11} = \Phi(\gamma_{11},N_{1},M). \label{eq:eps12}
\end{equation}
The user $u_2$ directly decodes its message, so the instantaneous BLER $\epsilon_{2}$ is
\begin{equation}
     \epsilon_{2} = \epsilon_{22} = \Phi(\gamma_{22},N_{2},M).  \label{eq:eps2}
\end{equation}
Then the average BLERs at the two users are obtained by
\begin{equation}
     \overline\epsilon_{1} = \mathbb{E}[\epsilon_1] \qquad \textrm{and} \qquad \overline\epsilon_{2} = \mathbb{E}[\epsilon_2].  \label{eq:epsaves}
\end{equation}

By taking the expectation of the instantaneous BLER over the SINR distribution average BLER $\overline\epsilon_{ij}$ is given as
\begin{equation}
    \overline\epsilon_{ij} = \int_{0}^{\infty} \Phi(\gamma_{ij},N_{j},M) f_{\gamma_{ij}}(x)dx
\end{equation}
\begin{equation}
    \approx \int_{0}^{\infty} Q \left(\frac{log_2(1+\gamma_{ij})-\frac{N_{j}}{M}}{\sqrt{\frac{v_{ij}}{M}}}  \right) f_{\gamma_{ij}}(x)dx, \label{eq:qfuncunt}
\end{equation}
where $f_{\gamma_{ij}}(x)$ is the probability density function (PDF) of the SINR $\gamma_{ij}$. Equation \eqref{eq:qfuncunt}, does not have a closed form solution and based on work the by Makki et al.\cite{Zorzi:2014}, 
$Q \left(\frac{log_2(1+\gamma_{i})-\frac{N_{i}}{M}}{\sqrt{\frac{v_{i}}{M}}}  \right) \approx \Xi_{i}({\gamma _{i}}) $ can be approximated as

\begin{equation} {\Xi_{i}}({\gamma _{i}}) = \begin{cases} 
    {1,}&{{\gamma _{i}} \leq {\upsilon _{i}},} 

    \\ {\frac{1}{2} - {\lambda_{i}}({\gamma_{i}} - {\theta_{i}}),}&{{\upsilon _{i}} < {\gamma _{i}} < {\tau_{i}},} 
    \\ {0,}&{{\gamma _{i}} \geq {\tau_{i}},} 
\end{cases} \label{eq:qapprox}\end{equation}
where \begin{align}
    {\lambda_{i}} &= \sqrt{\frac{M}{2\pi\left(2^{\frac{2 N_i}{M}}-1\right)}}, \qquad \theta_{i} = 2^{\frac{ N_i}{M}}-1,\\
    \upsilon _{i} &=\theta_{i} - \frac{1}{2\lambda_{i}} \qquad \textrm{and} \qquad  \tau_{i} =\theta_{i} + \frac{1}{2\lambda_{i}}.
\end{align}

Using this approximation in  \eqref{eq:qfuncunt}, the average BLER $\overline\epsilon_{i}$ is given by
\begin{align}
    \overline\epsilon_{i} = \lambda_{i} \int_{\upsilon _{i}}^{\tau_{i}} F_{\gamma_{i}}(x) dx \label{eq:epsCDF},
\end{align}
where $F_{\gamma_{i}}(x)$ is the cumulative distribution function (CDF) of the SINR $\gamma_{i}$.
\subsection{Average BLER for Decoding Far User's Information}
Based on the work by Cai et al.\cite{Cai:2018}, the CDF of the SINR $\gamma_{i2}$ for decoding of $u_2$'s message with HARQ-CC is derived as
\begin{multline}
    F_{\gamma_{i2}}(r) \approx c_i^T\sum_{\{p_1,\cdots,p_N\}\in \mathbb{P}}\Lambda \left[ \prod_{n=1}^{N}\Psi^{p_n}(a_n) \right]\\
     \times \sum_{k=1}^L (\omega_k \ln{2} ) E_1\left(\frac{S_{k,N}}{r}\right). \label{eq:CDFgammai2}
\end{multline}

The description of the variables and functions is given under \eqref{aveblerfinali2}.The proof is provided in Appendix \ref{app:epsCDF} as an extension of the work in \cite{Cai:2018}.

With the CDF of $\gamma_{i2}$ in \eqref{eq:CDFgammai2}, an approximation for the average BLER, $\overline\epsilon_{i2}$  can be computed using  \eqref{eq:epsCDF} as

\begin{subequations}\label{aveblerfinali2}
\begin{align}
      \overline\epsilon_{i2} &\approx  \lambda_{2} c_i^T\sum_{\{p_1,\cdots,p_N\}\in \mathbb{P}}\Lambda \left[ \prod_{n=1}^{N}\Psi^{p_n}(a_n) \right] \nonumber\\ 
       & \qquad \quad \times \sum_{k=1}^L (\omega_k \ln{2} ) [\Omega(\upsilon_{2},S_{k,N}) - \Omega(\tau_{2},S_{k,N})]\label{eq:epsi2final}
     \intertext{where,}
     c_i &= \frac{2\pi\kappa\alpha_2}{N \mu_i \rho} e^{\frac{1}{\mu_i \rho \alpha_1}},\quad \kappa = \frac{\alpha_2}{\alpha_1}, \quad \mu_i = \frac{1}{1+{d_i}^{\eta}},\\
     a_n &= \quad cos\left(\frac{2n-1}{2N} \pi\right) \> \textrm{for} \> \,n = 1,2,\cdots N,
\end{align}
    \begin{align}
    \Lambda &=\frac{T!}{\prod_{n=1}^Np_n!}, \quad
    \mathbb{P} =\left\lbrace p_1,...,p_N|T=\sum _{n=1}^Np_n \right\rbrace,\\
    & S_{k,N} = \frac{k \kappa \ln{2} }{2} \sum_{n=1}^N p_n (a_n+1), \label{eq:Skn}\\
    \Psi(a_n) & = \frac{\sqrt{1-a_n^2}}{(2\alpha_2-\alpha_1\kappa(a_n+1))^2}e^{-\frac{2\alpha_2}{\mu_i\rho \alpha_1 (2\alpha_2-\alpha_1\kappa(a_n+1))}},\\
    \omega_k  &= (-1)^{\frac{L}{2}+k} \sum_{\left \lfloor{j=\frac{k+1}{2}}\right \rfloor }^{\min\left(k,\frac{L}{2}\right)} \frac{j^{\left(\frac{L}{2}+1\right)}}{(\frac{L}{2})!} \binom{\frac{L}{2}}{j}\binom{2j}{j}\binom{j}{k-j},\\
    & \Omega(x,y) = x e^{-\frac{y}{x}}-(x+y)E_1 \left(\frac{y}{x}\right), \label{eq:phixy}
    \end{align}     
\end{subequations}
$E_1(x) = \int_x^\infty \frac{e^{-t}}{t} dt$ is the exponential integral function and $N,L$ are complexity-accuracy trade-off parameters. The proof is provided in Appendix \ref{app:aveblerfinali2}.

\subsection{Average BLER for Interference-free Decoding of the Near User's Information } \label{subsec:u1_int_free_NOMA}

For $u_1$ decoding its information with HARQ-CC after $T$ transmissions, the SNR is given by \eqref{eq:SNRu11} which is

\begin{align}
    Z & = \sum_{t=1}^{T}\gamma^{t}_{11}\nonumber = \sum_{t=1}^{T} \rho\alpha_1|\tilde h_{1,t}|^2 = \sum_{t=1}^{T} \rho\alpha_1 \mu_1 |h_{1,t}|^2,
\end{align}
where $ \mu_1 = \frac{1}{\sqrt{1+{d_1}^{\eta}}}$.
Since $h_{1,t} \sim \mathcal{CN}(0,1)$, $| h_{1,t}|^2$ is an exponential variable, $\rho\alpha_1 \mu_1 |h_{1,t}|^2$ is exponentially distributed such that $|h_{1,t}|^2 \sim Exp(\frac{1}{\rho\alpha_1 \mu_1 })$. The sum of $T$ exponential random variables is a Gamma distributed random variable with $T$ degrees of freedom. Therefore $Z$ can be described as

\begin{equation}
       Z \sim  Gamma (T,\frac{1}{\rho\alpha_1 \mu_1 }). 
\end{equation}
Then the CDF of $Z$ is,
\begin{equation}
    F_Z(r) = \frac{1}{\Gamma (T)} \gamma\left(T, \frac{r}{\rho\alpha_1 \mu_1 }\right) \label{eq:SNR11cdf}
\end{equation}
where $ \Gamma (k)=\int _{0}^{\infty }t^{k-1}e^{-t} dt $ is the Gamma function and ${\gamma (k,x)=\int _{0}^{x}t^{k-1}\,\mathrm {e} ^{-t}\, dt}$ is the lower incomplete Gamma function.

Therefore, $\overline\epsilon_{11}$  can be computed using \eqref{eq:epsCDF} resulting in
\begin{subequations}\label{eq:eps11bler}
\begin{align}
    \overline \epsilon_{11} = \lambda_{1} ( \Upsilon(\tau_{1}) - \Upsilon(\upsilon_{1})), 
\end{align}
where,
\begin{align}
    \Upsilon(x) &=  \frac{1}{\Gamma (T)} \left[ \gamma\left(T, \frac{x}{\rho\alpha_1 \mu_1 }\right)- \rho\alpha_1 \mu_1\gamma\left(T+1, \frac{x}{\rho\alpha_1 \mu_1 }\right)\right], \label{eq:upsilon}
\end{align}
\end{subequations}
 and the $\mu_1 , \lambda_1 , \tau_1 \> \textrm{and} \> \nu_1$ as defined before. The proof is provided in Appendix \ref{app:eps11bler}.
 
If the users are served using OMA, the blocklength or the number of channels uses available for transmission, $M$ would be shared between the two users and their messages will be transmitted utilizing the full power for that particular number of channel uses without interference from the other user. Note that the average BLER for OMA will have the same form as in \eqref{eq:eps11bler} with $\alpha_1 = 1$ and $\mu_1$ will be replaced by $\mu_i$ for $i =  1,2$.
 
\section{Blocklength and Power Allocation} \label{sec:blocklength_and_power}

\subsection{Asymptotic BLER approximations} \label{sec:asymptoticBLER}

Due to the mathematical complexity of the derived expressions in Section \ref{sec:NOMA_approx}, asymptotic expressions are derived in high SNR conditions. In short packet communications, the rate $\frac{N_i}{M}$ is small\cite{Xinyu:2019}, which leads to $\tau_{i,M}-\upsilon_{i,M}$ being smaller. Thus, the integration in (\ref{eq:epsCDF}) can be approximated using the Riemann integral approximation, $\int_a^bg(x)dx = (b-a)g(\frac{a+b}{2})$ such that
\begin{equation}
    \overline\epsilon_{i}^\infty \approx \lambda_{i}(\tau_{i}-\upsilon_{i}) F_{\gamma_{i}}(\frac{\tau_{i}+\upsilon_{i}}{2}) = F{\gamma_{i}}(\theta_{i}), \label{eq:epsireimann}
\end{equation}
where the superscript $\infty$ denotes the asymptotic approximation.

The average BLER targets for ultra reliable communication are in the order of $10^{-5}$ or lower and can be achieved with high transmit SNR. Therefore, $1-\epsilon_{12}^\infty \approx 1$ in \eqref{eq:epsaves}, which results in $\epsilon_1^\infty \approx \epsilon_{12}^\infty + \epsilon_{11}^\infty \label{eq:eps1approx}$. Therefore, $\overline\epsilon_1^\infty$ approximates to  $\mathbb{E}[\epsilon_{12}^\infty]+\mathbb{E}[\epsilon_{11}^\infty] = \overline\epsilon_{12}^\infty + \overline\epsilon_{11}^\infty$ and $\overline\epsilon_2^\infty$ can be obtained by  $\mathbb{E}[\epsilon_{2}^\infty] =\mathbb{E}[\epsilon_{22}^\infty] = \overline\epsilon_{22}^\infty$.

\subsection{Required blocklength and power allocation}
The problem of finding the required blocklength $M$, which guarantees the target BLERs can be stated as \begin{subequations}
    \begin{align}
    \textrm{find}  & \quad  M  \\
    \textrm{s.t} &\quad \overline\epsilon_1 \> = \> \overline\epsilon_1^{R} \label{eq:e1target} \\
    & \quad \overline\epsilon_2 \> = \> \overline\epsilon_2^{R} \label{eq:e2target}\\
    & \quad \alpha_1 + \alpha_2 = 1  \label{eq:alphas}\\
    & \quad 0 < \alpha_1 < 0.5  \label{eq:alpha1range}\\
    \textrm{for given} &\quad \rho,\mu_1,\mu_2, N_1, N_2, T,
    \end{align}\label{eq:problem}
\end{subequations} where the required BLERs for the two users are $\overline\epsilon_{1}^{R}$ and $\overline\epsilon_{2}^{R}$. The conditions in (\ref{eq:e1target}) and (\ref{eq:e2target}) ensure the reliability targets of the users while (\ref{eq:alphas}) and (\ref{eq:alpha1range}) arise from the NOMA principle. Since $\alpha_2 = 1- \alpha_1$, $\alpha_2$ can be omitted from the expressions.

According to Section \ref{sec:asymptoticBLER}, $\overline\epsilon_1^R$ and $\overline\epsilon_2^R$ can be expressed as,
\begin{align}
    \overline\epsilon_1^{R} = \overline\epsilon_{12}^\infty + \overline\epsilon_{11}^\infty \quad \textrm{and} \quad  \overline\epsilon_2^{R} = \overline\epsilon_{22}^\infty. \label{eq:targetblers}
\end{align}
Since $u_1$ is the stronger user with high channel gain and according to NOMA principle more power is allocated to $u_2$, average BLER for decoding $u_2$'s information in the first stage of SIC at $u_1$ is smaller than the average BLER for the second stage of SIC when interference-free decoding of $u_1$ is done. Therefore, for simplicity, the average BLER for the first stage of SIC in $u_1$ is considered as $\overline\epsilon_{12}^\infty = \delta \overline\epsilon_{11}^\infty$, where $\delta < 1$. 
Then, from (\ref{eq:targetblers}) $\overline\epsilon_{11}^\infty$ can be written as
\begin{equation}
    \overline\epsilon_{11}^\infty = \frac{\overline\epsilon_1^{R}}{1+\delta}.  \label{eq:epsR}
\end{equation}

Using the approximation with the Riemann integral as in  \eqref{eq:epsireimann}, $\overline\epsilon_{11}^\infty$ and $\overline\epsilon_{22}^\infty$ can be obtained as
\begin{align}
    \overline\epsilon_{11}^\infty & \approx  F_{\gamma_{11}}(\theta_{1}) \label{eq:reimann_approx_11}\\
    \overline\epsilon_{22}^\infty & \approx  F_{\gamma_{22}}(\theta_{2}) \label{eq:reimann_approx_22}.
\end{align}

Therefore, from \eqref{eq:reimann_approx_11} the blocklength $M$, which satisfies the reliability targets can be found as
\begin{equation}
    M = \frac{N_1}{log_2\left(1+\mu_1\rho\alpha_1\Gamma(T)\gamma^{-1}(T,\frac{\overline\epsilon_1^{R}}{1+\delta})\right)}, \label{eq:M1exp}
\end{equation}
where $\gamma^{-1}(k,x)$ is the inverse of the lower incomplete Gamma function. With the use of \eqref{eq:M1exp} in \eqref{eq:reimann_approx_22} the required $M$ can be found. Also, the required blocklength for OMA can be obtained by the addition of the blocklengths needed to achieve their reliability targets using a similar expression to \eqref{eq:M1exp}, with $\alpha_1 = 1$ and $\mu_1$  replaced by $\mu_i$ for $i =  1,2$.

Let $G$ be a function such that $ G(\alpha_1) \triangleq \overline\epsilon_2 \> - \> \overline\epsilon_2^{R}$ according to the  condition in (\ref{eq:e2target}). Solving $G(\alpha_1) = 0$ will give the $\alpha_1$ needed to achieve for the required blocklength, which can be used to find the required blocklength $M_{req}$ using (\ref{eq:M1exp}). Noting that $G(\alpha_1)$ is highly nonlinear, the solution can be computed using Algorithm 1.
\begin{algorithm}
1: \textbf{Input} : $\overline\epsilon_1^R, \overline\epsilon_2^R, \rho, \mu_1, \mu_2 ,N_1, N_2, T, \delta $ and tolerance $\nu$\\
2: \textbf{Output} : $M_{req} \> \textrm{and} \> \alpha_1^*$ \\
3: \textbf{Initialize} : $\alpha_1^- = 0$ and  $\alpha_1^+ = 0.5$ \\
4: \textbf{while} $|G(\alpha_1^c)| > \nu$ \textbf{do} \\
5:  set $\alpha_1^c \leftarrow (\alpha_1^+ + \alpha_1^-) / 2$ \\
6:  compute $G(\alpha_1^c)$ based on (\ref{eq:reimann_approx_22}) and (\ref{eq:M1exp})\\
7: \textbf{if} : $G(\alpha_1^c)G(\alpha_1^+) > 0$ \textbf{then} set $\alpha_1^+ \leftarrow \alpha_1^c$\\
8: \textbf{else} : set $\alpha_1^- \leftarrow \alpha_1^c$\\
9: \textbf{end while}\\
10: set $\alpha_1^* \leftarrow \alpha_1^c$ \\
11: compute $M_{req}$ using (\ref{eq:M1exp}) with $\alpha_1^*$
\caption{Power Allocation and Required Blocklength for NOMA with HARQ CC}
\end{algorithm}

\section{Numerical Results} \label{sec:numerical_results}
Monte Carlo simulations are carried out based on the results for the decoding error probability in short blocklengths to verify the accuracy of the approximations derived in Section \ref{sec:NOMA_approx}. In all the simulations, the complexity-accuracy parameters $N = 30$ and $L = 18$ to ensure the numerical accuracy. The path loss exponent $\eta = 2$ while $d_1 = 3\>m$ and $d_2 = 7\>m$.
\begin{figure}[ht]
  \begin{center}
    \includegraphics*[width=0.9\columnwidth]{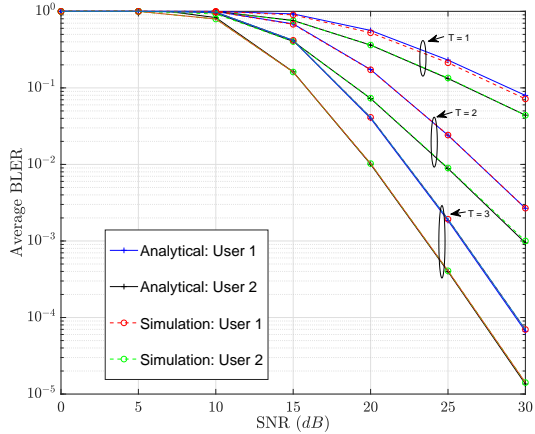}
  \end{center}
  \caption{Average BLER vs. transmit SNR ($\rho$) for different number of transmission rounds ($T$) with  $\alpha_1 = 0.1$, $\alpha_2 = 0.9$, $N_1 = N_2 = 160$ and $M = 200$.}
  \label{fig:bler_sim}
\end{figure}

Figure \ref{fig:bler_sim} shows the average BLERs plotted against the transmit SNR ($\rho$) for different maximum transmission rounds. The approximations derived match with the Monte Carlo simulation results, which prove the accuracy of the expressions in (\ref{aveblerfinali2}) and (\ref{eq:eps11bler}). According to Figure \ref{fig:bler_sim}, the far user always has a smaller average BLER than the near user, $u_1$. The reason is that higher power is allocated for the far user for user fairness in the NOMA principle. Also, with the increasing number of maximum transmission rounds allowed, the average BLER decreases for a particular transmit SNR.

\begin{figure}[ht]
  \begin{center}
    \includegraphics*[width=0.9\columnwidth]{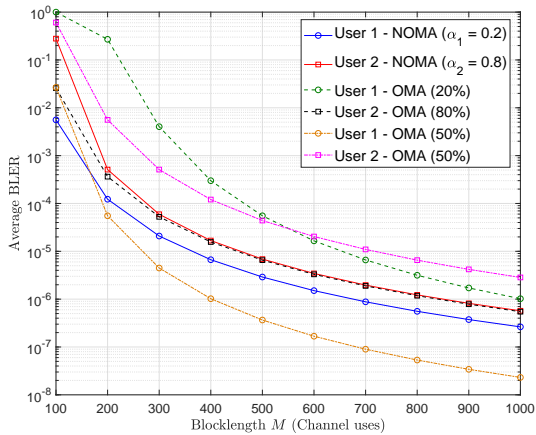}
  \end{center}
  \caption{Average BLER vs. blocklength for NOMA with $\alpha_1 = 0.2$ and OMA with $20\%$ and $50\%$ for $u_1$, $\rho = 30$ dB, $T = 3$ , $N_1 = N_2 = 300$.}
  \label{fig:blervsblocklength}
\end{figure}

In Figure \ref{fig:blervsblocklength}, average BLER is plotted with the blocklength at $\rho = 30$ dB, $T= 3$ with power allocation $\alpha_1$ set to 0.2 and $N_1$ and $N_2$ set to 300. The comparison with OMA is provided for two scenarios as 20\%-80\%  and 50\%-50\% blocklength share for $u_1,u_2$ respectively. It is clear from Figure \ref{fig:blervsblocklength}, when the blocklength increases the average BLERs in all scenarios decrease monotonically which is desirable. One interesting result is that the performance of $u_2$ in NOMA and OMA with 80\% share is almost similar with increasing blocklength. However, $u_1$ has a lower average BLER when NOMA is used compared to OMA with $20\%$ share of blocklength. Nevertheless, as the blocklength increase, this difference in performance between NOMA and OMA decreases. For the second scenario, blocklength is shared equally between the two users. The performance of $u_2$ degrades significantly compared to the performance with $80\%$ share. However, $u_1$ achieves a lower BLER than NOMA since a higher number of channel uses is available to $u_1$. Although $u_1$ has a lower average BLER with an equal share in OMA than NOMA, $u_2$'s average BLER degrades significantly. Therefore, the NOMA scheme delivers fairness to both users, unlike OMA, since the difference in average BLER between two users is smaller than in OMA while achieving considerable average BLER performance for both users.
\begin{figure}[ht]
  \begin{center}
    \includegraphics*[width=0.8\columnwidth]{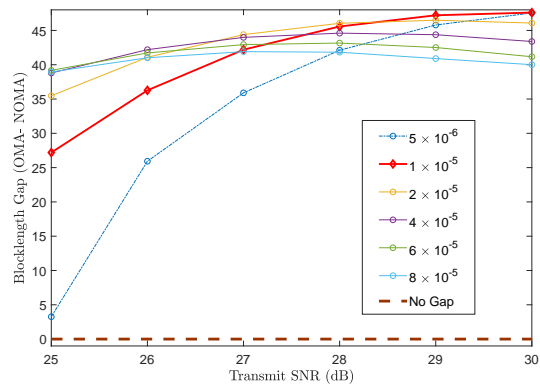}
  \end{center}
  \caption{Blocklength gap between OMA and NOMA vs. transmit SNR for $\overline\epsilon_1^R = 1 \times10^{-5}$ and varying $\overline\epsilon_2^R$ with   
   $N_1=N_2=300$, $T =3$, $\delta =0.1.$}
  \label{fig:blocklength_gap}
\end{figure}

Figure \ref{fig:blocklength_gap} shows the gap between OMA and NOMA for the required blocklength to achieve a given reliability target of $1 \times10^{-5}$ for $u_1$ and varying $\overline\epsilon_2^R$  for $u_2$ using Algorithm 1. The value of $\delta$ is set to $0.1$. Here the gap is taken by subtracting the NOMA blocklength from the OMA blocklength. It can be seen from Figure \ref{fig:blocklength_gap}, NOMA has a smaller blocklength than OMA for the given reliability targets since the gap is positive. The bold red curve represents both users having the same reliability target of $1\times 10^{-5}$ and NOMA always has a lower blocklength requirement and this gap increases as the transmit SNR increases. For lower reliability target such as $5\times 10 ^{-6}$ for $u_2$, the gap is smaller as seen from the dashed curve. Thus, NOMA has a lower blocklength requirement than OMA which leads to having lower latency when the reliability targets are in the order of $10^{-5}$. 

\section{Conclusion}  \label{sec:conclusion}
This paper analyzed the performance of NOMA with HARQ-CC for finite blocklength by deriving tight closed-form approximations for the average BLER for two users. The comparison with OMA was done proving that NOMA could meet lower average BLER requirements such as $1\times 10^{-5}$ while ensuring user fairness better than OMA. Further, an algorithm to determine the blocklength required to meet the reliability requirements of the two users was developed based upon the asymptotic expressions considering high SNR conditions. Simulations proved that NOMA has a lower blocklength requirement in high SNR leading to lower latency compared to OMA when the reliability requirements are in the order of $10^{-5}$. Analysis with multiple antennas is intended to be done as future work.
\appendices \label{sec:appendices}
\section{Proof of Equation \eqref{eq:epsCDF}} \label{app:epsCDF}
The PDF of the SINR for decoding $u_2$’s signal at $u_i$ where $i= 1,2$ after $T$ rounds of transmissions,  $Z = \gamma_{i2}  = \sum_{t=1}^{T}\gamma^{t}_{i2}$ is given by Equation (35) in \cite{Cai:2018}.  The CDF is calculated by extending that work. By taking the integral over $f_Z(z)$ with respect to $z$ as
\begin{subequations}
\begin{align}
    F_Z(r) &=  \int_{-\infty}^r f_Z(z) dz =  \int_{0}^r f_Z(z) dz. \nonumber\\
            & \approx c_i^T\sum_{\{p_1,\cdots,p_N\}\in \mathbb{P}}\Lambda \left[ \prod_{n=1}^{N}\Psi^{p_n}(a_n) \right] 
    \sum_{k=1}^L (\omega_k \ln{2} ) I_k,  \nonumber 
\end{align}
\begin{align}
    \textrm{where} \> \> I_k = \int_{0}^r \frac{1}{z} e^{\frac{-k \kappa \ln{2} }{2z} \sum_{n=1}^N p_n (a_n+1)} dz.  \nonumber 
\end{align}
\end{subequations}
By change of variables with $u = \frac{1}{z}$ integral in $I_k$ converts to,
\begin{align}
     I_k & = \int_{\frac{S_{k,N}}{r}}^\infty \frac{1}{u} e^{-u}du  = \> E_1\left(\frac{S_{k,N}}{r}\right)\nonumber
\end{align}
where $S_{k,N}$ as defined in \eqref{eq:Skn} and $E_1(x)$ is the exponential integral function defined by $E_1(x) = \int_x^\infty \frac{e^{-t}}{t} dt$. 
\section{Proof of Equation \eqref{aveblerfinali2}}\label{app:aveblerfinali2}
\noindent CDF of $\gamma_{i2}$ is given by \eqref{eq:CDFgammai2}. Computation of the approximation for average BLER, $\overline\epsilon_{i2}$ is provided here using \eqref{eq:epsCDF}.
\begin{subequations}
\begin{align}
    \overline\epsilon_{i2}  & \approx \lambda_{2} \int_{\upsilon_{2}}^{\tau_{2}}  c_i^T\sum_{\{p_1,\cdots,p_N\}\in \mathbb{P}}\Lambda \left[ \prod_{n=1}^{N}\Psi^{p_n}(a_n) \right] \nonumber\\
      & \qquad \qquad \times \sum_{k=1}^L (\omega_k \ln{2} ) E_1\left(\frac{S_{k,N}}{x}\right)dx \nonumber \\
     & = \lambda_{2} c_i^T\sum_{\{p_1,\cdots,p_N\}\in \mathbb{P}}\Lambda \left[ \prod_{n=1}^{N}\Psi^{p_n}(a_n) \right] 
      \sum_{k=1}^L (\omega_k \ln{2} ) J_k, \nonumber \\
     & \textrm{where} \>\> J_k = \int_{\upsilon_{2}}^{\tau_{2}} E_1\left(\frac{S_{k,N}}{x}\right) dx = -S_{k,N}\int_{\frac{S_{k,N}}{\upsilon_{2}}}^{{\frac{S_{k,N}}{\tau_{2}}}} \frac{E_1\left(v\right)}{v^2} dv.  \nonumber 
\end{align}
\end{subequations}
\begin{align}
\intertext{Using integration by parts twice and Leibniz integral rule}\nonumber
& = -S_{k,N}\left(\left[-\frac{E_1(v)}{v}+ \frac{e^{-v}}{v}\right]_{\frac{S_{k,N}}{\upsilon_{2}}}^{\frac{S_{k,N}}{\tau_{2}}}+ \int_{\frac{S_{k,N}}{\upsilon_{2}}}^{{\frac{S_{k,N}}{\tau_{2}}}}\frac{e^{-v}}{v} dv\right)\nonumber\\
& = \Omega(\upsilon_{2},{S_{k,N}}) - \Omega(\tau_{2},{S_{k,N}}), \nonumber 
\end{align}
where $\Omega(x,y)$ as defined \eqref{eq:phixy} which completes the proof.
\section{Proof of Equation \eqref{eq:eps11bler}} \label{app:eps11bler}
\noindent The average BLER, $\overline\epsilon_{11}$ can be computed using \eqref{eq:epsCDF} with the CDF for $\gamma_{11}$ given by the \eqref{eq:SNR11cdf} as
\begin{align}
    \overline\epsilon_{11} & = \lambda_{1} \int_{\upsilon_{1}}^{\tau_{1}} F_{\gamma_{11}}(x) dx  = \lambda_{1} \int_{\upsilon_{1}}^{\tau_{1}}\frac{1}{\Gamma (T)} \gamma\left(T, \frac{x}{\rho\alpha_1 \mu_1 }\right) dx. \nonumber \\
     \intertext{Using integration by parts,}
     & = \lambda_{1}\frac{1}{\Gamma (T)} \left( \left[ x \gamma\left(T, \frac{x}{\rho\alpha_1 \mu_1 }\right)\right]_{\upsilon_{1}}^{\tau_{1}} - Q_1 \right), \label{eq:Q1int} \\
     \intertext{ where $Q_1 = \int_{\upsilon_{1}}^{\tau_{1}} x \frac{d}{dx} \gamma\left(T, \frac{x}{\rho\alpha_1 \mu_1 }\right) dx.$ By definition of the lower incomplete Gamma function, change of variables with $u = \delta_1 x$ where $\delta_1 =\frac{1}{\rho\alpha_1 \mu_1}$ and Leibniz  rule }
    Q_1& = \frac{1}{\delta_1}\int_{\delta_1\upsilon_{1}}^{\delta_1 \tau_{1}} u \frac{d}{du} \left(\int_0^u t^{T-1} e^{-t} dt\right) du. \nonumber \\
    & = {\rho\alpha_1 \mu_1} \left[\gamma\left(T+1,{\frac{x}{\rho\alpha_1 \mu_1}}\right)\right]_{\upsilon_{1}}^{\tau_{1}}. \nonumber
\end{align}
Using the result of $Q_1$ in \eqref{eq:Q1int} completes the proof.
\bibliographystyle{di}
\bibliography{di}
\end{document}